\documentclass[12pt]{article}
\begin{document}
\def\R{{\hbox{{\rm I}\kern -.2em\hbox{\rm R}}}}
\def\cP{{\cal P}}
\def\cW{{\cal W}}
\def\cS{{\cal S}}
\def\cte{{\mbox{\rm c}^{\mbox{\rm te}}\; }}
\def\real{{\mbox{\rm I\hspace{-.2 em}R}}}
\def\arcsinh{\mathop{\rm arcsinh}\nolimits}
\def\arccosh{\mathop{\rm arccosh}\nolimits}
\begin{titlepage}
\begin{flushright} UMH-MG-98/02\\ ULB-TH-98/11 
\end{flushright}
\vskip 0.5 cm
\begin{center} {\LARGE\bf (2+1) dimensional stars}\\
\vskip 0.5 cm   M. Lubo\footnote{E-mail lubo@sun1. umh. ac. be}$\mbox{}^{\mbox{\footnotesize ,\ a}}$, 
M. Rooman\footnote{Ma\^\i tre de Recherches F. N. R. S. }$\mbox{}^{\mbox{\footnotesize ,\
}}$\footnote{E-mail mrooman@ulb. ac. be}$\mbox{}^{\mbox{\footnotesize ,\ b}}$,
Ph. Spindel\footnote{E-mail spindel@sun1. umh. ac. be}$\mbox{}^{\mbox{\footnotesize ,\ a}}$ \\
\vskip 0.25 cm
a) {\em M\'ecanique et Gravitation, Universit\'e de Mons-Hainaut}, 
\\ {\em 6, avenue du Champ de Mars, B-7000 Mons, Belgium}
 \\
\vskip 0.25 cm
b) {\em Service de Physique th\'eorique, Universit\'e Libre de Bruxelles}, \\ 
{\em  C. P. 225, bvd du Triomphe, B-1050 Bruxelles, Belgium}
\vskip 0.5 cm
\end{center}
\begin{abstract}
We investigate, in the framework of (2+1) dimensional gravity, stationary, rotationally symmetric
gravitational sources of the perfect fluid type, embedded in a space of arbitrary cosmological 
constant. We show that the matching conditions between the interior and exterior geometries imply
restrictions on the physical
 parameters of the solutions. In particular, imposing finite  sources and  absence of closed timelike
curves privileges negative values of the cosmological constant, yielding  exterior vacuum geometries
of rotating black
 hole type. In the special case of static sources, we prove the complete integrability of the field
equations and show  that the sources' masses
 are bounded from above and, for vanishing cosmological constant, generally equal to one. We also
discuss and illustrate the stationary configurations by explicitly solving the field equations for
constant mass--energy densities. If the pressure vanishes, we recover as interior geometries G\"odel
like metrics defined on causally well behaved domains, but with unphysical values of the mass to
angular momentum ratio. The introduction of pressure in the sources cures the latter problem and
leads to physically more relevant models. \end{abstract}
\vfill 
\end{titlepage}
\section {Introduction}
They are several, good or not so good, reasons to pay attention to gravity in lower dimensions. In
particular, as quoted by R. Jackiw \cite{Ja}, systems in a hot phase are phenomenologically described
in a manifold of topology $\Sigma \times S^1$ instead of $\Sigma  \times \R $; in the limit of
infinite temperature the circle $S^1$ shrinks into a point and we are left with an Euclidean theory
on $\Sigma$. Here we shall however not consider Euclidean, but only Lorentzian (2+1) solutions. Yet,
the latter may be relevant \cite{Ja} for the description of large 1--dimensional structures that seem
to be observed in the Universe, such as strings and vortices, whose interactions are governed by
(2+1)--gravity as far as we may ignore their extension in the third spacelike dimension.

On the other hand, pure gravity in (2+1) dimensions is not as trivial as it seems at first sight. It
has globally defined degrees of freedom whose physics is far from being totally understood (see for
example  ref. \cite{DGW}) and may thus possess some relevant features of (3+1)--gravity.  It seems
therefore worthwhile to improve our understanding of the classical physics it defines before tackling
the (3+1) problem, all the more because (2+1)--gravity is undoubtedly simpler
 than (3+1) 
\cite{GAK,CF1,RM,Mann}. In absence
of matter (vacuum solutions), the (2+1) geometry is locally de Sitterian, anti--de Sitterian or
Minkowskian. Globally however, things are more complicated. The full space--time appears, in general,
as the quotient of a covering space and a discrete isometry group. For instance, as shown in
\cite{BHTZ}, well--chosen identifications in anti--de Sitter (AdS) space yield the exterior geometry
of rotating black holes in (2+1) dimensions.

The present work was originally motivated by the finding \cite{RS} of a one parameter family of (2+1)--dimensional G\"odel like
metrics containing  the AdS geometry. This suggested the possibility of truncating the former and matching it to the latter,
so as to obtain  non vacuum solutions of (2+1)--gravity without causal pathologies. A preliminary investigation has shown that
such solutions indeed exist, and has led to a more systematic search for interior solutions  connected to a space
describing the exterior of black holes. These solutions are the (2+1) analogues of stars (and will be so called hereafter),
as they correspond to finite extended objects whose geometry matches an appropriate exterior geometry. 

In section 2, we establish the field equations and junction conditions,  assuming an abelian two parameter symmetry group of
the metric, a perfect fluid as gravitational source and (anti) de Sitter exterior geometries.
 In section 3, we first consider
static configurations and show that, due to the simplicity of the geometry and the
physics in (2+1) dimensions, they can be solved by quadratures, locally for positive values of the cosmological constant,
and globally  otherwise, extending previous works \cite{GAK,CF1,CF2,CZ}.  In section 4, we
consider rotating sources.  For constant mass--energy density, we were able to obtain analytic
solutions (contrary to what happens in (3+1) dimensions, where, to our knowledge, no analytical
solutions for compact objects have been obtained). One class of interior solutions contains
the aforementioned one parameter family of G\"odel like geometries, the other being
expressible in terms of elliptic functions. All the obtained solutions are causally well
behaved; some are nevertheless physically unacceptable, as their angular momentum is too large
compared to their mass. Indeed, such solutions would lead to naked causal singularities in
case of collapse, unless centrifugal forces prohibit them to evolve into black holes. This
point requires studying the singularity theorems in (2+1) dimensions, which is beyond the goal
of this paper. Section 5 presents a few concluding words.

\section{Stellar structure equations}

We assume the gravitational field equation to be 
\begin{equation} 
G_{\mu \nu}+\Lambda g_{\mu \nu}= \pi \; T_{\mu \nu} \quad , \label{Eeq}
\end{equation}
where we have conventionally fixed the gravitational coupling constant equal to $\pi$ and introduced a cosmological constant
$\Lambda$. The
metrics we shall consider will be supposed stationary and rotation invariant, i.e.  admitting a 2--dimensional abelian isometry
group. This means that in adapted coordinates, the metric components will depend only on a single variable. 
The interior geometry of the star is driven by the matter energy momentum tensor, assumed to be that of a perfect fluid~:
\begin{equation}
T_{\mu \nu}=(\sigma + p)\;u_{\mu}u_{\nu}+p\;g_{\mu \nu} \quad , \label{pf}
\end {equation}
where $\sigma$ is the mass--energy density of the fluid constituting the star,  $p$ the pressure, and
$\bf u$ the 3-velocity of the fluid, satisfying ${\bf u . u}= -1$. Moreover, we shall also suppose that $\sigma$ and $p$ are 
 positive definite and related
by an equation of state~:
\begin{equation}
\sigma=\sigma(p)\qquad,\qquad \sigma\geq 0\qquad,\qquad p\geq 0 \qquad . \label{etat}
\end {equation}

The interior metrics we shall consider hereafter take the general form~:
\begin{equation}ds^2_{in} = -  T[r]^2 dt^2 + B[r]^2 dr^2 + (Y[r]^2-Z[r]^2) d \phi ^2 
-2 T[r] Z[r] dt d\phi  \quad , \label{genmet}
\end{equation}  
where the functions $T[r]$,  $B[r]$, and $Y[r]$ are assumed to be non--negative. The function $B[r]$ is a gauge function that
 can be fixed at our convenience, at least locally.  If we assume this metric to possess a symmetry axis (located on $r=\bar 
r$) without conical singularities, we have to impose the conditions \cite{KSHM}~:
\begin{equation}
\lim_{r \rightarrow \bar  r} {{Y[r]^2 \over B[r]^2\,(r-\bar  r)^2} =1}  \qquad , \qquad  
\lim_{r \rightarrow \bar  r} {{Z[r] \over B[r]^2\,(r-\bar  r)^2} =\omega }  \qquad  , \label{reg}
\end{equation} 
where the constant $\omega$ is the angular velocity of the star near its center of rotation.

The 
most general expression of the metric for the vacuum exterior geometry has been established in 
\cite{BHTZ}  (for $\Lambda\neq
0$). It reads~:
\begin{equation}ds_{ex}^2= - [(N^{\perp}) ^2 - \rho ^2 (N^\theta )^2 ] d \tau ^2 + { 1 \over (N^{\perp})^2 } d
\rho ^2 + \rho ^2 d\theta ^2 + 2 \rho ^2 N^\theta  d\tau d \theta  \qquad , \label{exds2}
\end{equation} 
with 
\begin{eqnarray} 
N^\theta &=& - { J \over 2 \rho ^2 } + L_{\infty}  \qquad , \\
(N^{\perp}) ^2 &=&  -M - \Lambda\, \rho ^2 + {J^2 \over 4 \rho^2} \qquad , 
\end{eqnarray} where $M$ is the mass, $J$ the angular momentum, and  $L_{\infty}$ an
integration constant (denoted by $N^\phi(\infty)$ in ref. \cite{BHTZ}). This metric describes, for $\Lambda <0$,
(2+1)--dimensional black holes. Note that, as stressed in \cite{BHTZ}, $\rho=0$ does not correspond to a line in the geometry
(\ref{exds2}), but to a 2--dimensional surface, a cylinder whose circular sections are lightlike, on which the Killing vector
$\partial_\theta$ becomes null but not zero. This explains why this metric does not present a conical singularity, despite the
fact that the conditions (\ref{reg}) are not satisfied at  $\rho=0$.  

The matching conditions on the surface, separating the regions where  the internal and external
geometries are defined, impose the equality of the intrinsic geometry and of the extrinsic curvature with respect to both
geometries \cite{LI}. For all the solutions we shall discuss, the equation of these surfaces will be of constant
radial coordinate $r=r_{\star}$ in terms of coordinates $\{t,r,\phi\}$ covering the interior region. Such surfaces are orbits
of the 2 parameter isometry group, and their equations are thus also simply $\rho=\rho_{\star}$ in terms of the exterior
coordinates.  On the junction surface, the $\{t,r\}$ and
$\{\tau,\rho\}$ coordinates are related by linear transformations, whereas the angular variables $\phi$ and
$\theta$ may be identified, as they are both assumed to vary  between $0$ and $2\pi$. Hence, the
coordinate transformation relations on the connection surface read~: 
\begin{equation}\phi=\theta\qquad,\qquad \tau = T_0 \;  t  \qquad \mbox{\rm and}\qquad \rho = R_0 (  r+r_0) \qquad ,
\label{coordtrans}
\end{equation} 
with $T_0$, $R_0$ and $r_0$ constants to be determined. The sign of $T_0$ will be supposed positive, the time $t$ and $\tau$
flowing in the same direction. The sign of $R_0$ will indicate what part ($\rho > \rho_{\star}$ or $\rho < \rho_{\star}$) of the
so--called exterior geometry will be glued at the junction surface. If $R_0<0$, we have to take the part $\rho < \rho_{\star}$
and  obtain a closed space, presenting generically a singularity at $\rho =0$, hidden or naked. In the following we shall
limit ourselves to the cases
$R_0>0$, except when stated otherwise. Note furthermore that we have  used the space--time isometry group to fix at zero the
 arbitrary additive constants in the two first equalities;  $r_0$  becomes meaningful
 once the remaining gauge freedom in the internal metric is removed by completely specifying the internal $r$ coordinate.

Owing to the  transformations (\ref{coordtrans}), the equality of the induced metric on the junction surface implies the continuity
of the interior and exterior metric components expressed  in the same coordinates~:
\begin{equation}
g^{in}_{\mu\nu}[r_\star]=g^{ex}_{\mu\nu}[\rho_{\star}]\qquad,\qquad \mu,\nu \in \{t,r,\phi\}\qquad.\label{cong}
\end{equation} 
The equality of the extrinsic curvature with respect to the two space--time geometries reduces, in the particular case
considered here (junction surface at $r=r_\star$), to require the continuity of some of the metric component
derivatives~:
\begin{equation}\left. \partial_r g^{in}_{\mu\nu}[r]\right |_{r=r_\star}=\left. \partial_r g^{ex}_{\mu\nu}[\rho[r]]\right
|_{r=r_\star}
\qquad,\qquad \mu,\nu  \in \{t,\phi\}\qquad.\label{condg}
\end{equation} 
Conditions (\ref{cong}, \ref{condg}) imply seven
equations that must be satisfied on the surface of the star $r=r_{\star}$~:
\begin{eqnarray} 
T^2_{\star}=\left[-M-\Lambda \rho_{\star}^2+JL_{\infty}-L_{\infty}^2\rho_{\star}^2\right]T_0^2&
&,\label{match1}\\ 
\frac {T_{\star}T_{\star}'}{R_0}=-(\Lambda + L^2_{\infty})\rho_{\star}
T_0^2& &,\label{match2}\\
B_\star^2=\frac{R_0^2}{-M-\Lambda \rho_{\star} ^2+\frac{J^2}{4\rho_{\star}^2}}& &,\label{match3}\\
{Y^2_{\star}-Z^2_{\star}}=\rho_{\star}^2& &,\label{match4}\\
\frac 1{R_0}\left(Y_{\star}Y'_{\star}-Z_{\star}
Z'_{\star}\right)= \rho_{\star} & &,\label{match5} \\
 T_{\star}Z_{\star}= \left(\frac J2
-\rho_{\star}^2 L_{\infty}\right)T_0 & &,\label{match6}\\ 
\frac 1{R_0}\left(T'_{\star}Z_{\star}+T_{\star}Z'_{\star}\right)= -2 \rho_{\star} L_{\infty}T_0 & &
.\label{match7} \end{eqnarray}
These seven equations fix the six unknowns $\{L_{\infty},\rho_\star,T_0,R_0,J,M\}$~:
\begin{eqnarray} 
L_{\infty}&=&-\left.   \frac{(T\,Z)'}{2\,B\,  T\, Y}  \; \right  \vert_{r=r_\star}\qquad,\label{Ninf}\\
\rho_\star&=&\left.(Y^2-Z^2)^{1/2}\right\vert_{r=r_\star}\qquad,\label{rhostar}\\
T_0&=&\left. \frac{2\,B\, T\, Y}{(Y^2-Z^2)'}  \,  \right\vert_{r=r_\star} \qquad,\label{T0}\\
R_0&=&\left. \frac{(Y^2-Z^2)'}{2\,(Y^2-Z^2)^{1/2}}\right\vert_{r=r_\star} \qquad,\label{R0}\\
J&=&\left.   \left(\frac{Y^2-Z^2}{T Z}\right)'\frac{T\,Z^2}{B\,Y}  \; \right\vert_{r=r_\star} \qquad,
\label{J}\\
M&=&- \frac {R_0^2}{B^2} - \Lambda \rho_\star^2 + \frac {J^2}{4 \; \rho_\star^2} \qquad,\label{M}
\end{eqnarray}
where we have used the positivity assumptions of $R_0$ and $T_0$.  We obtain moreover  the additional consistency relation~:
\begin{equation}
\left. \frac{T'Y'}{B^2 TY}+\frac14 \left[\frac T{BY} \left(\frac ZT\right)'\right]^2\right\vert_{r=r_\star}+\Lambda=0\qquad .
\label{consist}
\end{equation}

In order to solve the Einstein equations (\ref{Eeq}), we found useful to introduce the triad~: 
\begin{equation}
\theta^0=T[r]\, dt + Z[r]\, d \phi \quad, \quad 
\theta^1=B[r]\,dr \quad, \quad
\theta^2= Y[r]\, d \phi \quad, \label{triad}
\end{equation} 
such that~:
\begin{equation}
ds^2_{in} = - ( \theta^0 )^2  + ( \theta^1 )^2 + ( \theta^2 )^2 \qquad . 
\end{equation} 
With respect to the basis (\ref{triad}), the Einstein tensor has 4 non-trivially
vanishing components : 
\begin{eqnarray} 
G_0^0 &=& { \frac1{B\,Y}\left(\frac{Y'}{B}\right)'} - { 3 \over 4} \left[ \frac T{B\,Y} \left( \frac ZT \right) ' \right]^2 \qquad
,\label{TenE0}\\ 
G_1^1 &=& \frac {T' Y'} {B^2T\, Y} + { 1 \over 4} \left[ \frac T{B\,Y} \left( \frac ZT \right) ' \right]^2 
\qquad ,\label{TenE1}\\  
G_2^2 &=& { \frac1{B\,T}\left(\frac{T'}{B}\right)'} + { 1 \over 4} \left[ \frac T{B\,Y} \left( \frac ZT \right) ' \right]^2
\qquad \label{TenE2},\\  
G_0^2 &=&{ 1 \over {2\,B\, T^2}} \left[ \frac {T^3}{B\,Y} \left( \frac ZT \right) ' \right]'
\qquad . \label{TenE02} 
\end{eqnarray} 
Our symmetry assumptions imply that the fluid's velocity $\bf u$ has to be a linear combination of the two Killing vector
fields. Accordingly, the corresponding one--form $\underline{\bf u}$ may be written~:
\begin{equation}
\underline{\bf u}=\cosh [v]\, \theta^0 + \sinh [v] \theta^2 \qquad , \label{u}
\end{equation}
where $v$ may depend on $r$.
Using the radial $G_1^1$--Einstein equation,
\begin{equation}
G_1^1= \pi \, p-\Lambda\qquad ,
\end{equation}
we see that the consistency equation (\ref{consist}) 
is satisfied if and only if the pressure vanishes at the surface of the star. This condition is a quite natural requirement
from a physical point of view. Indeed, a discontinuity of the pressure is incompatible with the hydrostatic equilibrium equation of
 the fluid (the Bianchi identity)~:
\begin{equation}
	p'=- \; \frac{p+\sigma}{TY}[ \cosh ^2 [v] T'Y - \sinh ^2 [v] TY' - \cosh [v] \sinh [v] T^2 \left ( \frac ZT \right ) '
] \qquad.
\label{BianchiG}
\end{equation}

In what follows, we shall  build explicit stellar models, under the simplifying ans\"atz that the fluid's velocity one--form
is aligned with
$\theta^0$, i.e. $v=0$, but not necessarily  integrable. We shall consider both
the static ($\underline{\bf u}$ integrable) and stationary ($\underline{\bf u}$ non integrable) cases.
Under this assumption, eq.(\ref{BianchiG}) simplifies into~:
\begin{equation}
p'=-\frac{T'}{T}(\sigma +p) \label{Bianchi} \qquad .
\end{equation}
This equation and the inequalities (\ref{etat}) imply that, if the star possesses a surface, $T'$ must be positive on it.
As a consequence, under the assumption of the positivity of $R_0$, we obtain the bound~:
\begin{equation}
L_{\infty}^2\leq -\Lambda \qquad , \label{Lneg}
\end{equation}
and thus $\Lambda \le 0$. This shows that finite perfect fluid structures of the type described here need an AdS or
Minkowskian exterior space, unless  $R_0<0$, in which case the surrounding space is closed. 

To close the general considerations of this section, let us remind that
in the AdS case, the metric (\ref{exds2}) presents black hole type horizons located at~:
\begin{equation}
\rho_{H\pm}^2= \frac M{-2\,\Lambda}\left(1\pm\sqrt{1+\frac{J^2\,\Lambda}{M^2}}\right) \qquad ,\label{hor}
\end{equation}
if $M>0$ and $-\Lambda\,J^2/M^2<1$.
It also  presents a surface of infinite redshift, an ergosphere, at~:
\begin{equation}
\rho_{erg}^2 =\frac{M - J\,L_\infty}{-\Lambda - L_\infty^2} \qquad. \label{erg}
\end{equation} 
Equation (\ref{match1}) shows that the radius of the junction surface is always in the static region, i.e. in a region of finite
redshift. The star models considered here can thus never present an ergosphere (nor a blackhole horizon), if $R_0$ is positive.
The origin of this constraint lies in our choice of the interior metric (\ref{genmet}), for which $g_{tt}$ is explicitly
negative\footnote{Lorentzian rotating solution with
$g_{tt}>0$ could be obtained by exchanging the definitions of $\theta^0$ with $\theta^2$ in eq. (\ref{triad}), but they
never present centers of rotation.}. 

\section{ Static stars}

The most general (2+1)--dimensional, static, rotation invariant metric can be
written as 
\begin{equation}
ds^2_{in}=-T^2[r]\,dt^2+B^2[r]\,dr^2+Y^2[r]\, d\phi^2\quad ,
\end{equation}
with the range of the  variable $\phi$  fixed to $[0,2\pi]$. The Einstein 
equations (\ref{Eeq})
become, taking this ans\"atz and eqs (\ref{TenE0}--\ref{TenE02}) into account, and with as matter source a perfect fluid at
rest [eqs (\ref{pf}\ and \ref{u}) with $v=0$]~:
\begin{eqnarray}
	G^{0}_{0}&\equiv & \frac 1{BY} \left(\frac {Y'}{B}\right)'=- \pi \sigma - 
	\Lambda\equiv - \cS \qquad , \label{G00}\\
	G^{1}_{1}&\equiv & \frac {T'Y'}{B^{2}TY} =  \pi p- 
	\Lambda\equiv  \cP \qquad , \label{G11}\\
		G^{2}_{2}&\equiv&\frac 1{BT} \left(\frac {T'}{B}\right)'   =  \pi p- 
	\Lambda\equiv  \cP \qquad , \label{G22}
\end{eqnarray}
with as initial conditions~:
\begin{equation}
T[0]=1 \quad , \quad B[0]=1 \quad , \quad Y[0]=0 \quad\mbox{\rm and} \quad Y'[0]=1\quad ,\label{init}
\end{equation} 
in order to avoid conical singularities (eq. \ref{reg}) and to maintain the central pressure $p[0]$ finite. 
By equating the lefthand sides of eqs (\ref{G11}, \ref{G22}) and using the initial
conditions (\ref{init}), we obtain~:
\begin{equation}
T' = \cP [0]\, B\, Y \qquad .  \label{eqAC}
\end{equation} 
 If the function $Y[r]$ is constant, $\cal S$ and $\cal P$ vanish
and the  space is flat. So, without loss of generality, we shall hereafter assume that $Y[r]$ is not
constant. Mimicking the Oppenheimer--Volkoff \cite{OV} integration of the stellar structure equations in (3+1) dimensions, we
may define a radial coordinate $Y[r]=r$. We immediately obtain from eqs (\ref{G00} and \ref{init})~:
\begin{equation}
	B^{-2}[r]=1-2 \pi \int^r_0  \sigma(x) x dx\ -\Lambda r^{2}\qquad . \label{B-2}
\end{equation}

The Bianchi identity (\ref{Bianchi}) becomes:
\begin{equation}
\frac {d\cP}{dr}=-r\cP (\cP + \cS ) B^{2}	\qquad . \label{hydro}
\end{equation}
 This equation shows that, unless $\Lambda=0$, we cannot have a static, pressureless star. 
 Indeed, a cloud of dust cannot remain static in an expanding universe, driven here by a cosmological
constant. 
Using eq. (\ref{G00}), written in the form 
\begin{equation}\frac {dB}{B}= r\, \cS \, B^{2}\,dr \qquad ,
\end{equation} 
we obtain from eq. (\ref{hydro})~:
\begin{equation}\left(\frac 1{\cP + \cS}-\frac1\cP\right)d\cP=\frac{dB}B \qquad . \label{eqPB}
\end{equation} 
In order to go ahead, we now use the equation of state (\ref{etat})
describing the model. It allows us to integrate eq. (\ref{eqPB}), which
gives~:
\begin{equation}\frac{B[r]\cP[r]}{W[r]}=\frac{\cP [0]}{W[0]} \qquad , \label{eqPBW}
\end{equation} 
where $W[r]=w[p[r]]$ and $w[p]$ is the index of the fluid (the thermodynamical temperature in case of
radiation) defined by~:
\begin{equation}w[p]=w[p_0]\exp\int^p_{p_0}\frac{dq}{\sigma(q)+q}\qquad . \label{eqw}
\end{equation} 
The $G^1_1$--Einstein equation (\ref{G11}) can be integrated in the same way~:
\begin{equation}\frac{dT}{T}=\cP\, B^2\, r\, dr=-\frac{d\cP}{\cP+\cS}=-\frac{dW}{W}  \qquad . \label{eqWT}
\end{equation} 
The solution of this equation yields, using (\ref{init}), the well known Tolman thermal equilibrium condition 
\cite{Tol}:
\begin{equation}
T[r]W[r]=W[0] \qquad . \label{eqAW}
\end{equation} 
Equations (\ref{eqPBW}, \ref{eqw} and \ref{eqAW}) give a parametric representation of the solution
as a function of a thermodynamical variable, let say the pressure. A geometrical
parametrisation, in terms of the radial coordinate $r$, is obtained from eqs (\ref{eqPBW} and \ref{eqWT}),
which combine into~:
\begin{equation}\frac{\cP\, d\cP}{(\cP + \cS)W^2}=\frac{\cP\, dW}{W^3}=
- \left(\frac{\cP [0]}{W[0]}\right)^2d\left(\frac{r^2}2 \right) \qquad , \label{eqdr2}
\end{equation} 
and furnish, after integration, the implicit dependence of $p$ as a function of~$r$. 

Let us emphasize that we have here only discussed the local integration of the interior field
equations. We have still to consider  their domain of validity and the matching of the internal
solutions with  appropriate (AdS, flat or de Sitter) external spaces.

\subsection{AdS-like stars} 

We  first consider the AdS case $\Lambda <0$,  already discussed in detail by Cruz and Zanelli (CZ)
\cite{CZ}.
 As $p\geq 0$, we have $\cP[r] >0$, and thus $Y'[r]\neq
0$.
 Accordingly, the choice of the $r$ variable is valid inside the whole star.  Moreover, eq.
(\ref{eqdr2}) shows that the radius of the star will be finite or infinite according to the behavior
of the index $w$ near $p=0$. If $\lim_{p\rightarrow 0} w[p] =0$, the star will have an infinite
radius. For instance, if the star is of pure radiation $(\sigma=2\,p,\ w\propto p^{1/3})$, we obtain,
following the previous scheme~:
\begin{eqnarray}
T[r]&=&\frac{p[0]^{1/3}}{p[r]^{1/3}}   \qquad ,\\
B[r]&=& \frac{p[r]^{1/3}}{p[0]^{1/3}} \; \frac { \pi p[0] - \Lambda}{ \pi p[r] - \Lambda}   \qquad ,
\end{eqnarray}
and after elementary integration~:
\begin{equation} 
\left(\frac{\Lambda+2 \pi p[0]}{p[0]^{2/3}}\right)-
\left(\frac{\Lambda+2 \pi p[r]}{p[r]^{2/3}}\right)=
\left(\frac{\pi p[0]-\Lambda}{p[0]^{1/3}}\right)^2 r^2 \qquad ,
\end{equation}
which reduces to a cubic equation for $p[r]^{1/3}$.
Otherwise, when $w[0]\neq 0$,  we shall denote by $r_{\star}$ the finite value of the radial
coordinate of the connection surface and by the same subscript ($\star$) the values that
the various functions ($\cP, T, B, Y, T', Y', \dots$) take on it. As  
$B$ must be  positive on $r_{\star}$, we obtain the CZ upper bound for the mass \cite{CZ}~:
\begin{equation}
m_{\star}\equiv 2\pi\int_0^{r_\star} \sigma(x) x dx \leq 1 - \Lambda r_{\star}^2 \qquad . \label{mass}
\end{equation} 

For a static star, the angular momentum $J$ and the integration constant $L_{\infty}$ obviously vanish and the exterior geometry
around the star is given by the metric~:
\begin{equation}
ds^2_{ex}=-(-M-\Lambda \rho^2)\,d\tau^2+\frac{d\rho^2}{-M-\Lambda \rho^2}+\rho^2 d\phi^2 \label{AdS}
\qquad .
\end{equation}  
The matching conditions (eqs \ref{Ninf}-\ref{M}) and eqs (\ref{B-2} and \ref{mass}) imply 
the mass relation~:
\begin{equation}
 M= m_{\star}-1 \label{eqM} \qquad , 
\end{equation} 
already given in ref. \cite{CZ}.\\
This equation illustrates the fact that the exterior mass parameter $M$ is always larger than $-1$ (under the assumption
$\sigma>0$), with the extreme value $M=-1$ corresponding to the usual, singularity--free, AdS space 
\cite{BHTZ}. However, we
would like to stress that we do not find a mass gap between $M=-1$ and $M\ge 0$, contrary to what happens for black holes.
Note also that eqs (\ref{mass} and \ref{eqM}) directly imply the absence of blackhole horizons for $M>0$, the horizon
$\rho_{H+}$ (eq. \ref{hor}) corresponding to a surface located inside the star~:   
\begin{equation}
\rho_{H +}^2= \frac M{-\Lambda} < \rho_{\star}^2 = r_{\star}^2 \qquad .
\end{equation} 

We see furthermore that an observer at rest on the surface of such a star (an observer whose world line is $\phi=Cst$ and
$r=r_{\star}$) follows an accelerated trajectory, with an invariant acceleration~:
\begin{equation}
a=\sqrt{a_{\alpha}a^{\alpha}}=\left\vert \frac{T'_\star}{B_\star T_\star}\right\vert=
\frac{ - \Lambda r_{\star}}{\sqrt{-M-\Lambda r_{\star}^2}} \qquad , 
\end{equation}  
directed in the direction of increasing $r$. It is interesting to 
notice that, while the gravitational interaction does not propagate in 
(2+1) dimensions, the acceleration of an observer at the surface of the 
star depends both on $\Lambda$ and $M$. More precisely, as the radial coordinate $r_{\star}$ has a
geometrical meaning, this equation illustrates the fact that the acceleration depends on the presence
or not of a star. Indeed, though gravity does not propagate, it manifests itself by  holonomy
effects \cite{Ja}, affecting here the definition of $r_{\star}$  as lengths of circles around the star. 

\subsection{de Sitter-like stars}

As shown at the end of section 2, finite objects embedded in a de Sitter environment require $R_0$ to be negative. This
implies that the full space is closed, the so--called exterior region being described by a metric like (\ref{AdS})
but with
$\Lambda >0$ and
$\rho\in [0,\rho_\star]$. In other words, there is no concept of interior or exterior regions for such spaces; all
constant
$t$--time sections are compact.

To discuss these models, we skip to the  gauge where $B =1$
and perform the change of radial coordinate~: $ \xi = \int B[r] dr $. Indeed,  the
function
$Y[\xi]$ is here not everywhere increasing inside the star and the previous parametrisation, $Y[\xi]=r$, remains only
locally  valid. In the $B=1$ gauge, the Einstein equations (\ref{G00} --\ref{G22}) become~:
\begin{eqnarray}
\frac{Y''}{Y} = - ( \pi \sigma + \Lambda )& =& - \cS \quad , \label{G002}\\
 \frac{ T'Y'}{TY} =   \pi p - \Lambda &=& + {\cal P}  \quad , \label{G112}\\
 \frac{T''}{T} =  \pi p - \Lambda  &= &+ {\cal P}  \quad , \label{G222}
\end{eqnarray}
 with the same initial conditions (\ref{init}). The first equation (\ref{G002}) shows that $Y[\xi]$ is a
convex function, bounded by
$ \Lambda^{(-1/2)}
\sin[\Lambda^{1/2}\, \xi]$ on its physical (i.e. $Y\geq 0$) interval of definition, which is included in the
interval
$[0,\pi\,\Lambda^{1/2}]$. This justifies a posteriori our gauge choice. Indeed, using eqs (\ref{eqAC} and
\ref{G112}), we see that
$Y[\xi]$ reaches its maximum when
${\cal P}=0$; its derivative vanishes and starts to become negative.  But  as $\Lambda > 0$, the pressure
is still positive and we have thus not yet reached the star's surface.

Once the solution $Y[\xi]$ of eq. (\ref{G002}) is obtained, we may integrate eq. (\ref{eqAC}) which gives~:
\begin{equation}
T[\xi]=1+\cP [0]\,\int_0^{\xi}Y(x)\,dx \qquad . \label{TY}
\end{equation}
This equation, introduced into the junction condition (\ref{match2}), yields~:
\begin{equation}
T'_\star T_\star=\cP [0]\,Y_\star T_\star=-\Lambda\,R_0 \, T_0^2\, \rho_\star\qquad ,
\end{equation}
implying $\cP [0]>0$. The function $T[\xi]$ is thus increasing.
Combining eqs  (\ref{Bianchi} and \ref{eqAC}), we now obtain~:
\begin{equation} 
{p}' = - {\cal P}[0] (\sigma + p)\frac{Y}{T} \qquad . \label{Bianchi2}
\end{equation}
The pressure hence decreases with $\xi$ and it depends on the specific form of the equation of state whether or not a
singularity occurs.

Solving the junction conditions (\ref{match1}--\ref{match5}), taking care of the sign of the radial derivative, which is
opposite to the one used in the previous case, we obtain~:
\begin{equation}
T_0=-\frac{T_\star}{Y'_\star}\quad,\quad R_0=-Y'_\star \quad,\quad \rho_\star=Y_\star \quad,\quad
M=-\Lambda Y^2_\star-Y^{\prime 2}_\star <0\quad. \label{decadix}
\end{equation}
This closed space possesses two centers of rotation. One is located at $\xi=0$, the other at $\rho=0$. By assumption (eq.
\ref{reg}),  there is no conical singularity on the first axis, but in general the second suffers from an angular defect of~:
\begin{equation}
\delta=(1-\sqrt{-M})2\,\pi \qquad .
\end{equation}

To be complete, note that if the initial conditions are such that $\cP [0]<0$, the function $T[\xi]$ is decreasing
and $p'$ is positive, so that 
$p$ cannot  vanish. We have thus generically a singularity there where $T[\xi]$ vanishes (see eq. \ref{eqAW}), except if
$Y[\xi]$ re--vanishes first, and moreover if  at this second zero 
$Y'[\xi]$ is equal to
$-1$, in which case we obtain a closed space with 2 antipodal centers of rotation and free of conical singularities. 

So we conclude that, in
general, there do not exist static stars  whose matter satisfies the phenomelogical conditions (\ref{pf}, \ref{etat}), in
a singularity--free (2+1)--dimensional space--time with positive cosmological constant.

By way of illustration, suppose the energy density $\sigma $ constant, so that $\cS$ is a positive constant. We easily obtain
the expression of the interior metric components  and the radial dependence of the pressure~:
\begin{eqnarray} 
 Y[r] &=& \frac{ \sin[ \sqrt{\cS} \; r]}{ \sqrt{\cS}}\qquad ,\\
 T[r] &=& 1+ \frac{{\cal P}[0] (1 - \cos[\sqrt{\cS}r])}{\cS} \qquad ,\\
p[r] &=& \frac{ p[0] + \sigma } {T[r]} - \sigma \qquad .
\end{eqnarray} 
The space admitting such a geometry appears as the product of a 2--sphere of radius $\cS ^{(-1/2)}$ with a line of time.
The surface of vanishing pressure is given by the solution of
\begin{equation}
\cos[{ \sqrt{\cS} r_{\star}}] = 1-\frac{\cS \, p[0]}{\sigma \,\cP [0]} \qquad ,
\end{equation}
which, according to the above discussion, requires $\cP [0] >0$ to exist. Using eq. (\ref{decadix}), we obtain for this model 
\begin{equation}
M=-1 + \frac { \pi \, \sigma}{ \pi \, \sigma + \Lambda} \sin^2 [\sqrt{\cS}r_\star] \qquad,
\end{equation}
which shows that the conical singularity is unavoidable.

\subsection{Minkowskian--like stars}

If $\Lambda = 0$, we may either use the gauge $B=1$ or choose $Y[\xi]=r$ to discuss the
global properties of the internal solutions. The function $Y[\xi]$ is indeed everywhere
increasing inside the star. Let us first assume that the pressure is not identically
zero; the maximum of $Y[\xi]$ then coincides with the vanishing of the pressure,
occurring just on the surface of the star. Using eqs (\ref{eqw},\ref{eqdr2}), it is easy 
to see that the radial coordinate $(r_\star)$ of the surface of the star is 
infinite when the
index of the fluid vanishes with the pressure as $p^{\alpha}$ with $\alpha > 1/2$. This
criterion generalizes the results obtained for  polytropic fluids in ref. \cite{CF1}.

As the curves $(\phi=Cst$, $r=r_\star)$ are
accelerated trajectories, it is impossible to match these interior geometries across a
surface $\rho=Cst$ to a flat space of metric~: 
\begin{equation}
ds^2_{ex} = - dt^2 + d\rho^2 + dy^2 \quad ,
\end{equation}
expressed in minkowskian
coordinates. 
The exterior metrics that continue the interior geometries are given by~:
\begin{equation}
ds^2_{ex} = - 
 \rho^2 d\tau^2 + d\rho^2 + dy^2\quad , \label{Rin}
\end{equation}
corresponding to flat geometries written in Rindler coordinates, in a (2+1)--dimensional Minkowski space whose
spacelike direction $y$ has been rendered periodic by identifying $y$ with $y + 2 \pi Y_{\star}$. This
periodicity condition results from the matching conditions for the
$g_{\phi\phi}$ component of the metrics,  while the continuity of their derivatives  reduces to $p_\star=0$. The junction
conditions  due to the
$g_{tt}$ metric components are
\begin{equation}      
T_{\star} = T_0 \rho_{\star} \qquad , \qquad
T'_{\star} = T_0\qquad .\label{eqAx}
\end{equation}

An unexpected result that emerges from the hydrostatic equilibrium equation
(\ref{hydro}) is that, assuming a (reasonable) equation of state,  a star with pressure and finite radius has always 
its mass parameter $m_{\star}$ equal to 1 if
$\Lambda=0$. Indeed,
if $B^{-2}[r]$ does not vanish with $p$ at the star's surface, Cauchy theorem applied to eq. (\ref{hydro}) implies that $p=0$
everywhere in the star. As a consequence, $B^{-2}_{\star}=0$ and, using eqs (\ref{B-2} and \ref{mass}), we get $m_\star =1$ and,
by virtue of eq.(\ref{eqM}), $M=0$. The metric (\ref{AdS}), with $\Lambda=0$, becomes thus singular, which confirms the
necessity of using the metric (\ref{Rin}). This universal value of the mass can also be directly  obtained from the
Einstein equation (\ref{G002}) as follows~:
\begin{equation}
    m_\star = \pi \int_{0}^{r_{*}^2}  \sigma dr^2      
 = \pi \int_{0}^{Y_{*}^2} \sigma  dY^2        
=- 2 \int_{0}^{Y_{*}}  Y'' dY     
 =-\int_{1}^{0} dY^{\prime 2}   =  1 \quad . \label{mass1}
\end{equation}
This result has already been  noticed in the special case
of a uniform density star in \cite{GAK} and for polytropic fluids in \cite{CF1}, and
demonstrated in a different way in  \cite{CF2}. But, contrary to what is claimed there, we
see from eq. (\ref{eqAx}) that the surface of the star follows an  accelerated trajectory with
respect to the exterior flat space. Physically, this is due to the fact that in absence of
gravitational interaction, a fluid shearing an internal pressure has to expand. Moreover, an
observer at the surface of such a star feels an acceleration directed towards the center of
the star, of magnitude $[ a = \xi_{\star}^{- 1} = \frac{T'_{\star}}{T_{\star} }]$. If this
observer\footnote{For more information about the life of inhabitants in a (2+1)--dimensional
world, see ref.  \cite{EAA}}  jumps from the surface of the star, he will follow an inertial
geodesic trajectory in the surrounding flat space. Nevertheless, he will fall again on the
accelerated surface of the star because he will be recaptured by it. This is actually also
what happens for observers in (3+1)--dimensional Schwarzchild metric except that in (2+1)
dimensions the escape velocity is the velocity of light.

If the star is pressureless (a special case also studied in ref. \cite{CF1}), the function
$T[\xi]$ is constant and $\sigma$ becomes an  arbitrary function of $\xi$. Indeed, as  there
is no gravitational interaction  between the particles of the the dust constituting the star,
they may be distributed with an arbitrary radial dependence. If the convex function $Y[\xi]$ is
such that at the surface of the star $Y'_{\star}$ is still non negative, the exterior solution
is given by the flat space geometry~: \begin{equation}
ds^2_{ex}=-(1-m_\star) \; d\tau^2+ \frac {d\rho^2}  {1-m_\star} + \rho^2 d\phi^2\qquad \mbox{\rm
with}\qquad
\rho\geq Y_{\star}\qquad .
\end{equation} 
This geometry presents  an angular deficit
$\delta =(1 -\sqrt{1 - m_{\star}})2\, \pi$, i.e. $m_\star = 1$ remains the maximal mass allowed for (a
circularly symmetric) object in this kind of universe, if we want to preserve its locally Minkowskian
character. Note that if $Y'_\star$ is negative, the interior space matches the cylindrical portion $\rho<Y_\star$ of
 a flat space, presenting a conical singularity on the axis $\rho=0$.

\section{ Rotating star}

When the star is uniformly rotating, its metric stops being static but remains stationary. In
the B=1 gauge, it can be written~:
\begin{equation}ds^2_{in} = -  T[r]^2 dt^2 + dr^2 + (Y[r]^2-Z[r]^2) d \phi ^2 
-2 T[r] Z[r] dt d\phi \qquad ,  \label{rotmet}
\end{equation}  
with the condition $Z[r] T[r] \not \equiv 0$. Under the assumption (\ref{u}) that the fluid 3-velocity is not tilted with respect
to the frame (\ref{triad}),  Einstein's equations  impose that $G_0^2$ (\ref{TenE02}) be equal to zero. This implies that $Y[r]$
is of the form~:
\begin{equation}
Y[r]= \frac 1c T[r]^3 \left( { Z[r] \over T[r] }\right) ' \qquad , 
\end{equation} 
where $c$ is a constant. When a center of rotation exists, eq. (\ref{reg}) yields~:
\begin{equation}
\omega=\frac c2 \qquad ,
\end{equation}
leading to interpret the constant $c$ as the angular velocity at the center of the star.
The remaining three
non-trivially satisfied Einstein equations are~:
\begin{eqnarray} 
G_0^0 &\equiv& - { { 3 c^2 } \over {4 T^4}}  + 3 {{T'} \over { T}} \left[ \frac { \left(T^2 (Z/T)' \right)'}
{T^2 (Z/T)'} \right] + 
 {{ T Z''' - T''' Z} \over {T Z'-  T' Z}} = - \pi \sigma - \Lambda  \quad ,  \label{ET0}\\
G_1^1 &\equiv& { {c^2} \over {4  T^4}} +{{T'} \over { T}} \left[ \frac { \left(T^3 (Z/T)' \right)'}
{T^3 (Z/T)'} \right]  =  \pi p - \Lambda  \quad , \label{ET1} \\
G_2^2 &\equiv& { {c^2 } \over {4 T^4}} +{{T'' } \over { T}} =  \pi p - \Lambda  \quad . \label{ET2} \\
\end{eqnarray} Our choice of the velocity field $\bf u$ imposes  the equality of $G_1^1$ and
$G_2^2$, from which we deduce that~:
\begin{equation}{{T'' } \over { T}} = {{T'} \over { T}} \left[ \frac { \left(T^3 (Z/T)' \right)'}
{T^3 (Z/T)'} \right] \qquad . \label{2SOL}
\end{equation} 
This is satisfied by two types of solutions,
\begin{eqnarray} 
T[r]&=&Cst \qquad \qquad \qquad \qquad \mbox{\rm or} \label{peq0}\\
Z[r]&=&\alpha\, T[r]+\beta\, T^{-1}[r]\qquad ,\label{pneq0}
\end{eqnarray}
where $\alpha$ and $\beta$ are constants. They are hereafter referred to as dust-like  and
pressurized (star) models, respectively.  These interior solutions can  match various
``exterior" spaces, but, like for the static case, only those of AdS type avoid 
singularities. We shall thus limit the subsequent analysis to $R_0$ positive and $\Lambda$
negative. Rotating perfect fluid solutions with $\Lambda = 0$, which exemplify the
occurrence of pathologies, are given in ref. \cite{CF3}.

\subsection{Dust--like models}

These solutions are characterized by eq. (\ref{peq0}) and hence, without loss of generality,  by
$T[r] = 1$. The pressure $p$ is thus constant, and we put it equal to zero in order that the star admits a
boundary. Let us emphasize that this model implies  $\Lambda =-c^2/4 < 0$ ; a negative
cosmological constant stabilizes the system by counterbalancing the centrifugal force. The
remaining function $Z[r]$ has to satisfy eq. (\ref{ET0}) which reduces to~:
\begin{equation}
Z'''+(\pi \sigma + 4\Lambda)Z'=0\quad . \label{ZG00}
\end{equation}
This means that  $Z[r]$ either defines the pressureless matter (dust)  repartition, or, conversely, is itself defined by the
matter repartition. If we assume the existence of a center of rotation at $r=0$, the regularity conditions (\ref{reg}) imply the
initial conditions for eq. (\ref{ZG00})~:
\begin{equation} 
Z[0]=0\qquad ,\qquad Z'[0]=0 \qquad , \qquad  Z''[0]=2\omega= c = 2\,\sqrt{-\Lambda}\qquad .
\end{equation}
As the pressure vanishes everywhere, the consistency condition (\ref{consist}) is identically satisfied by virtue of the Einstein
equations. The radius $r_{\star}$ of the surface of the star remains a free parameter.
Applying the general matching
conditions (eqs 
\ref{Ninf}--\ref{M}) to this
specific case, we find all parameters of the exterior metric as functions of $Z[r]$ and its first two derivative evaluated on
$r_{\star}$, except the integration constant which is constant~:
\begin{equation}
L_{\infty} = - \sqrt{-\Lambda} \qquad ,\label{Linf1}
\end{equation}
and reaches its upper bound allowed by eq. (\ref{Lneg}).
This equality implies, by virtue of eq. (\ref{match1}), the
additional condition~:
\begin{equation}
\frac { - \Lambda \; J^2  } {M^2}  >1 \qquad,
\end{equation}
when $M>0$. This relation severely limits the physical relevance of all dust--like star solutions. Indeed, in case of
collapse, a naked singularity occurs, whatever the sign of $M$ \cite{BHTZ}.

Nevertheless, we find it worthwhile to illustrate these models
in the special case $\sigma =Cst$, if only because we recover G\"odel like geometries analogous to those considered in
\cite{RS,RT,RTi,BD}. The function $Z$ is then obtained from the Einstein equation (\ref{ZG00})~: 
\begin{equation} 
Z[r]= \left\{ \begin{array}{lclr}
g\, \exp [r/a] + h\, \exp [-r/a] +k & \quad \mbox{\rm if}\quad & a^{-2}=-\pi \sigma - 4\Lambda
>0& ,\\ 
g\, \sin [r/a] + h\, \cos [r/a]  +k & \quad \mbox{\rm if}\quad &a^{-2}=+\pi \sigma + 4\Lambda >0 &  .
\end{array}
\right. \label{Z}
\end{equation}
 
Changing, if necessary, $r$ into $r+Cst$,
the metric (\ref{rotmet}) simplifies into four forms~:
\begin{eqnarray}
ds^2_{in}=&-&dt^2+dr^2
+\gamma^2\left\{
\begin{array}{l}
  \sinh^2[\frac ra]-a^2c^2(\cosh[\frac ra]+\kappa)^2   \\
  \exp[\frac {2r}a]-a^2c^2(\exp[\frac ra]+\kappa)^2    \\
  \cosh^2[\frac ra]-a^2c^2(\sinh[\frac ra]+\kappa)^2   \\
  \sin^2[\frac ra]-a^2c^2(\cos[\frac ra]+\kappa)^2   
\end{array}\right.\ d\phi^2 \nonumber \\
&-&2 a\;c\;\gamma\left\{
\begin{array}{l}
 \cosh[\frac ra]+\kappa \\
 \exp[\frac ra]+\kappa \\
 \sinh[\frac ra]+\kappa \\
 \cos[\frac ra]+\kappa 
\end{array}\right.
dt\;d\phi\qquad \qquad .\qquad \qquad \qquad \qquad\begin{array}{r}
(\ref{abcd}.{\rm a})\\
(\ref{abcd}.{\rm b})\\
(\ref{abcd}.{\rm c})\\
(\ref{abcd}.{\rm d})
\end{array}
\nonumber\\
&&\mbox{\hspace {20cm}}\label{abcd}
\end{eqnarray}
The first three metrics correspond to low density stars, i.e. $\pi \sigma <- 4\Lambda$, with the product $gh$ being 
$>0,\ =0$ or $<0$, respectively. The last metric corresponds to a high density star~: $\pi \sigma >- 4\Lambda$. Locally, the three
metrics labeled (a,b,c) are equivalent, as they have the same Einstein tensor  (while the
coordinate transformations\footnote{If we accept to consider complex coordinate transformations, then 
we may go from the  metric (\ref{abcd}.a) to (\ref{abcd}.c) by the following change~:
$r\mapsto r + i\frac {\pi}4 a$, $\phi \mapsto i\;\phi$ and $\kappa \mapsto i\; \kappa$. Such
transformation may be relevant in a semi-classical theory, where it could describe tunneling
transitions.}\ 
 that link them is not always obvious, see for instance \cite{RS,RTi}). But as the
interior solutions are given by restricting  the  domain of the $r$ variable to $r<r_{\star}$, they
correspond to different non-diffeomorphic subsets of a larger space on which $r$ is unrestricted. As
a consequence, the metrics (\ref{abcd}.{a,b,c}) have to be considered as distinct solutions.

It is easy to convince oneself that only the geometries (\ref{abcd}.{a}) and  (\ref{abcd}.{d}) can have
 a non-singular
center of rotation, if we fix~:
\begin{equation}\gamma=a\qquad,\qquad \kappa=-1\qquad ,
\end{equation} 
in which case, using eq. (\ref{reg}),  the angular velocity $\omega=c/2= \pm \sqrt{|\Lambda|}$. Setting 
\begin{equation}
\mu=a\; c=\left\{ 
\begin{array}{lcr}
(1-\frac {\pi\sigma}{4\,|\Lambda|})^{-1/2}\geq 1 \quad &\mbox{\rm if }&\quad \sigma \leq 4|\Lambda|/\pi \qquad ,\\
&&\\
(\frac {\pi\sigma}{4\,|\Lambda|}-1)^{-1/2}\geq 0 \quad &\mbox{\rm if }&\quad \sigma \geq 4|\Lambda|/\pi \qquad ,
\end{array}\right. \label{mu}
\end{equation} 
 we end up with two 1--parameter
families of metrics. The first family, corresponding to $\sigma \leq 4|\Lambda|/\pi$, is given by~:
\begin{eqnarray}
ds^2_{in}&=&-dt^2+ dr^2 + 4\; a^2\left(\sinh^2[\frac r{2a}]+(1-\mu^2)\sinh^4[\frac r{2a}]\right)d\phi^2\nonumber \\
&&-2\;a\; \left(2\;\mu\sinh^2[\frac r{2a}]\right)\;dt\;d\phi\qquad . \label{fam1}
\end{eqnarray}
This family has been studied in \cite{RS}, essentially from a geometrical point of view; it also appears as the non trivial part
of solutions of Einstein--Maxwell/Einstein--Maxwell--scalar field equations \cite{RT,RTi} and more recently in the framework of
a low energy string effective action \cite{BD}.  The elements of this family can be viewed \cite{RS} as squashed AdS
geometries~: for
$\mu=1
$, the metric describes a regular AdS space in unusual coordinates, whereas for
$\mu^2 =2$, it corresponds to the non trivial 3--dimensional part of the G\"odel metric \cite{Go}. All these geometries,
except the AdS one, are $SO(2,1)\times SO(2)$ invariant. Moreover, closed timelike curves pass through  each of their
points. In particular, the circles centred on the origin whose radii exceed the threshold value~:
\begin{equation}
r_c =2\,a\, \arcsinh \left[(\mu^2-1)^{-1/2}\right]=a \, \log \left[{ \mu +1 \over \mu -1}\right] \qquad ,
\end{equation}
are typical examples of such curves.
To avoid these causality pathologies, we have to fix the radius $r_{\star}$ of the surface of the star to be less then $r_c$.
Using eqs (\ref{J} and \ref{M}), we find that the mass and angular momentum parameters of the connected external AdS space are
expressed in terms of the parameter
$a$ and
$\mu$ as~: 
\begin{eqnarray}
J&=&
 - 4\, a \, \mu \,\left( {{\mu }^2}-1 \right) \,
     {{\sinh ^4[\frac r{2\,a}] }}\qquad ,\\
M&=&
-1 + 4\,(\mu^2-1) \sinh ^2[\frac r{2\,a}] -2\,(\mu^2-1)(\mu^2-2)\sinh ^4[\frac r{2\,a}]  \, .
\end{eqnarray}

The second 1--parameter family of metrics, which corresponds to $\sigma \geq 4|\Lambda|/\pi$, reads~:
\begin{eqnarray}
ds^2_{in}&=&-dt^2+ dr^2 + 4\; a^2\left(\sin^2[\frac r{2a}]-(1+\mu^2)\sin^4[\frac r{2a}]\right) d\phi^2\nonumber \\
&&+2\;a\; \left(2\;\mu\sin^2[\frac r{2a}]\right)\;dt\;d\phi \qquad . \label{fam2}
\end{eqnarray}
Due to the higher mass--energy density, the maximally analytic extensions of these spaces are topologically
$S^2\times
\real$ instead of $\real ^3$. These geometries also contain closed timelike curves. Again, to
avoid causality inconsistencies, the radius of the surface of the star has to be chosen less than a critical value~:
\begin{equation}
r_c = 2\,a \, \arcsin \left[(\mu^2 +1)^{(-1/2)}\right]\qquad .
\end{equation} 
The mass and angular momentum parameters of the connected external AdS space are~: 
\begin{eqnarray}
J&=& 4 \, a \, \mu \,\left( {{\mu }^2}+1 \right) \,
     {{\sin ^4[\frac r{2\,a}] }} \qquad , \\
M&=&-1 + 4\,(\mu^2+1) \sin ^2[\frac r{2\,a}] -2\, (\mu^2+1)(\mu^2+2)\sin ^4[\frac r{2\,a}] \, .
\end{eqnarray}

By way of illustration, plots of $|J|$ and $M$ as functions of $r_\star$, for the 2 families of metrics (\ref{fam1} and
\ref{fam2}) and for various values of
$\mu$ and thus of the mass--energy density
$\sigma$, are depicted in figs [1-2]. All curves are stopped at the limiting value~:
\begin{equation}
r_{lim}=\left\{
\begin{array}{lc}
(\mu /2)  \arccosh [\mu^2/(\mu^2-1)]&\mbox{\rm if}\quad \sigma \leq 4|\Lambda|/\pi \qquad ,\\
&\\
\mu \arcsin\left[1/\sqrt{2(\mu^2+1)}\right]&\mbox{\rm if}\quad \sigma \geq 4|\Lambda|/\pi \qquad ,\label{rlim}
\end{array} \right.
\end{equation}
beyond which $R_0$ becomes negative.

\subsection{Pressurized models}

The other types of solutions to eq. (\ref{2SOL}) are given by eq. (\ref{pneq0}) and are characterized by $p \neq  0$.
If the regularity conditions
(\ref{reg}) at the origin are imposed, we obtain furthermore that~: 
\begin{equation}
\alpha=-\beta \qquad, \qquad T'[0]=0 \qquad, \qquad c/2= \alpha\, T''[0]=\omega\qquad .\label{alpha}
\end{equation} 
As T[0] may set equal to 1, the metric reads~:
\begin{equation}
ds^2_{in}= - T^2 dt^2 + dr^2 - 2\,\alpha (T^2 - 1) d\phi dt + \alpha ^2\left[ \frac{T'^2}{\omega^2}  - (T - {1
\over T})^2 \right ] d\phi ^2\quad .
\end{equation} 
In this case, the  matching condition (\ref{Ninf}) yields~:
\begin{equation}
L_\infty=-\omega \qquad ,
\end{equation}
with 
\begin{equation}
\omega^2 < -\Lambda \qquad ,  \label{wmax}
\end{equation}
according to eq. (\ref{Lneg}).

Let us again illustrate this class of solutions by considering the special case of constant  mass--energy density. By
integrating twice the Einstein equation (\ref{ET0}), using the expression of $Z[r]$ and eq. (\ref{ET2}) to fix an integration
constant, we obtain~: 
\begin{equation} 
T'^2 = {\omega^2} ( {1 \over T^2} -1 ) + 2 \, \pi (p[0] + \sigma ) (T-1) - ( \pi \, \sigma +
\Lambda) (T^2 -1) \quad ,
\end{equation} 
showing that the explicit solution could be expressed in terms of elliptic functions. More interesting is the relation
resulting from the elimination of $T''[r]$ in eq. (\ref{ET2})~:
\begin{equation}
T= { p[0] + \sigma \over p[r] + \sigma } \qquad ,
\end{equation}
which allows to express the matching parameters in terms of the mass--energy density $\sigma$, the central pressure $p[0]$ and
the central angular velocity $\omega$,   $\alpha$ being obtained from eq. (\ref{alpha})~:
\begin{equation}
\alpha=\frac \omega{(-\Lambda + \pi \, p[0] - \omega^2)} \qquad .
\end{equation}
We get~:
\begin{eqnarray}
J&=& -\frac{2\;\pi p[0]^2 \, \omega}{\sigma (\omega^2+\Lambda - \pi p[0])^2} \qquad ,\label{J1}\\
M&=&\frac{ \pi\, p[0]^2  (-\Lambda +\omega^2) \; -\sigma (\omega^2+ \Lambda)^2}{\sigma \; (\omega^2 +\Lambda -  \pi \,
p[0])^2}\qquad .\label{M1}
\end{eqnarray}
We leave to the reader the check that the limit of $M$ for vanishing $\omega$ can also be obtained by
applying the method exposed in section 3 to the special case of constant mass--energy density.

The causality requirement
$Y^2_{\star}-Z^2_{\star}>0$ restricts the range of the variable
$\omega$ to~:
\begin{equation}
\omega^2<-\Lambda +\frac{\pi \, \sigma \, p[0]}{p[0]+2\sigma}\qquad ,
\end{equation}
which is automatically satisfied owing to the inequality (\ref{wmax}). 
From the latter, it is easy to verify that $M$ is always larger than $-1$ for positive $\sigma$ and
$p[0]$. However,  it  becomes positive only if the central pressure is high enough~:
\begin{equation}
p[0]^2>\frac{\sigma\,(\Lambda +\omega ^2)^2}{\pi\,(-\Lambda +\omega^2)} \qquad .\label{Mzero}
\end{equation}
Moreover, for $M>0$, the requirement $\sqrt{-\Lambda}|J|/M\leq 1$ imposes~:
\begin{equation}
p[0]^2 > \frac {\sigma(\Lambda+\omega^2)^2}{\pi (\sqrt{-\Lambda} -|\omega | )^2} \qquad ,\label{JoverM}
\end{equation}
which implies condition (\ref{Mzero}). This establishes the existence of physically acceptable solutions, for
sufficiently large values of the pressure.  Finally, the condition ensuring the absence of
tachyonic matter, $\sigma > p$ 
\cite{HE}, imposes a lower bound for the mass--energy density~:
\begin{equation}
\sigma_{min} = \frac {(\Lambda+\omega^2)^2}{\pi (\sqrt{-\Lambda} -|\omega | )^2} \qquad ,\label{smin}
\end{equation}
to avoid any causality pathology.

The above discussion is illustrated in fig. [3] where we have plot, in a ($\sigma\ ,p^2[0]$) plane, 
 the boundary curves  $\sigma=p[0]$,
$M=0$ and
$-\Lambda\;(J/M)^2=1$ for $\omega = 0.1 \, \sqrt{-\Lambda}$. These curves delimit the physically allowed values of the
mass--energy density and central pressure. 

\section{Conclusion}

This incursion in (2+1)--dimensional gravity confirms the qualitative difference between the roles played by positive and
negative cosmological constants. We have indeed shown that relevant elementary star models, 
obeying the physical criteria that space--time be free of naked
singularities and regions of causality violations, require
$\Lambda<0$. We have also seen that, while the hydrostatic equilibrium equation (see eq. \ref{Bianchi}) is similar to the one
we obtain in (3+1) dimensions, the gravitational potential
$T'/T$ it involves manifests itself only  trough the cosmological expansion or contraction of the space
and holonomy effects, in accord with
our understanding of (2+1)--gravity. The interpretation of the kinematics of the static star embedded in flat space is
particularly illustrative from this point of view. It inflates, its surface follows a curve a constant acceleration and its
mass is a universal constant. The fact that such star models with ``time" independent mass--energy density do exist, 
results from a subtle cancellation between volume expansion and Lorentz contraction. Indeed, the equal ``time" planes are
not the usual Minkowskian $t=Cst$ sections, but the boosted Rindler time sections.

When the surrounding space is AdS, the physics is richer. The
cosmological constant acts as an attractive gravitational force, increasing with the radial distance, equilibrated by the
mechanical effects of the pressure and the centrifugal forces. But pressure plays another, paradoxal
 role. As  seen in section 4, it also acts as a ``gravitational source", which counterbalances the centrifugal
effect. It was indeed shown that taking  pressure into account allows to obtain physically acceptable solutions, i.e.
singularity free solutions without closed timelike curves, which, in case of subsequent collapse, lead to black holes without
naked singularities. The main reason is that, owing to the pressure, the angular velocity of the star may be less than its
upper bound (see eq.
\ref{wmax}). The deep significance of the pressure's dual behavior is far from being completely clear (for us).
In (3+1) dimensions, we know that the pressure contributes, together with the mass--energy density, to the gravitational
attraction. In contrast, the Newtonian limit of (2+1)--dimensional Einstein gravity is a theory in which only the pressure is
the source of the potential \cite{GAK}, thereby enlightening its special role.
However, the relative signs between the cosmological constant, the squared angular velocity, the pressure
and the mass--energy density render difficult to draw from eqs (\ref{J1}, \ref{M1}) a complete 
physical intuition, similar to the one that has been built  from (3+1)--dimensional experiences. These
differences must be kept in mind before extrapolating to (3+1) dimensions physical conclusions from
the (toy) model that (2+1)--gravity theory is. 

Note that we have focused here on circular star models admitting a
center. However, other models that do not satisfy the regularity conditions
(\ref{reg}) are surely interesting to consider. They may indeed present geometrical pecularities  relevant to
illustrate various aspects of wormholes, ergospheres, etc. They will be considered elsewhere.

\strut \\
{\bf Acknowledgments}
We are very grateful to Marc  Henneaux for enlightening comments and 
discussions. We also thank Neil Cornish, Robert Mann and the referee 
for drawing our attention on various works on the subject.

\newpage
{\bf{Figure captions}}\\
\begin{itemize}
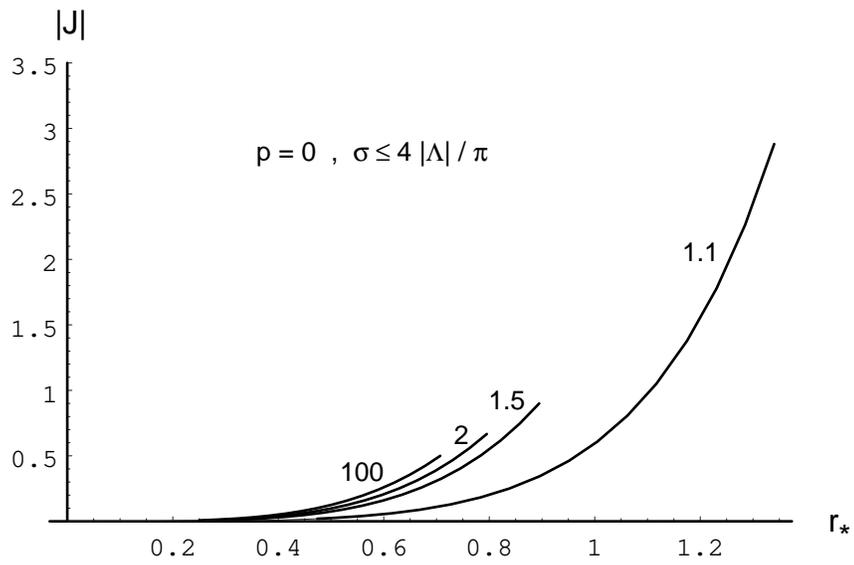
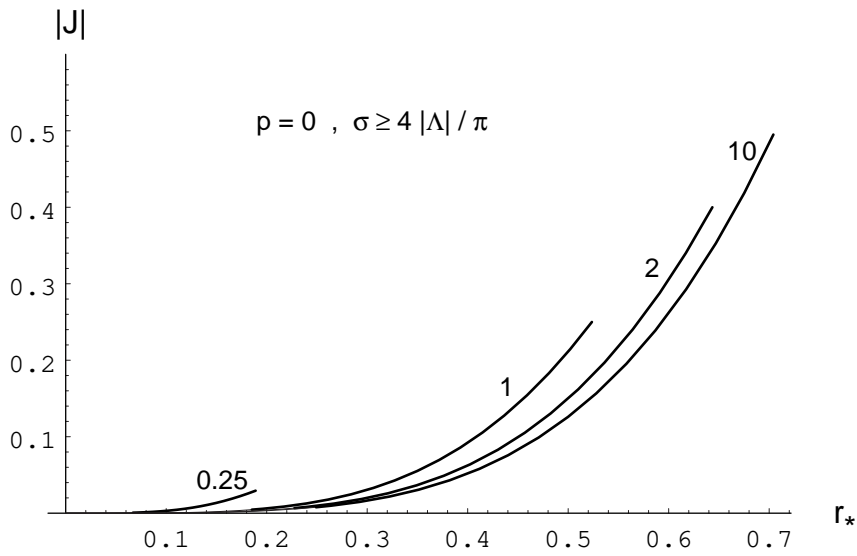
\item{Figure 1.} Plots of the asymptotic angular momentum $|J|$ (in units $|\Lambda| =1$), for constant mass--energy density
dust--like star models, as a function of the radial surface coordinate 
$r_\star$, for various values of $\mu$~:\\
{\bf a)} $\mu = 1.1,\, 1.5,\, 2,\, 100$ and $\sigma \leq 4\,|\Lambda|/\pi$;\\
{\bf b)} $\mu =  .25,\, 1,\, 2,\, 10$ and $\sigma \geq 4\,|\Lambda|/\pi$.\\
All curves are stopped at the critical
radius
$r_{lim}$ given in  eq. (\ref{rlim}).
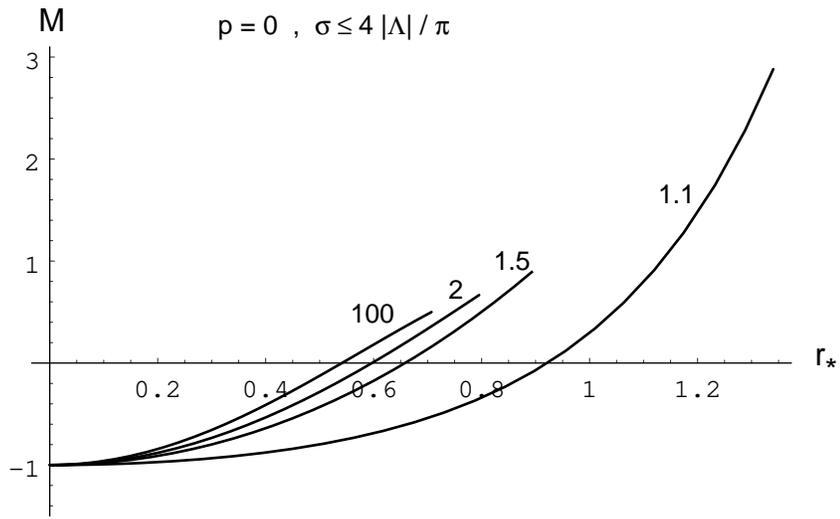
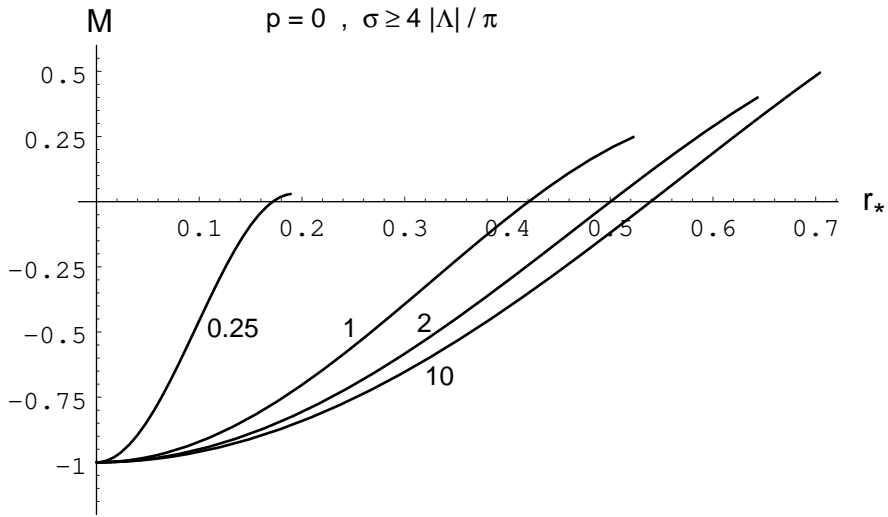
\item{Figure 2.} Plots of the asymptotic mass $M$ (in units $|\Lambda| =1$), for constant mass--energy density
dust--like star models, as a function of the radial surface coordinate 
$r_\star$, for various values of $\mu$~:\\
{\bf a)} $\mu = 1.1,\, 1.5,\, 2,\, 100$ and $\sigma \leq 4\,|\Lambda|/\pi$;\\
{\bf b)} $\mu = .25,\, 1,\, 2,\, 10$ and $\sigma \geq 4\,|\Lambda|/\pi$.\\
All curves are stopped at the critical
radius
$r_{lim}$ given in  eq. (\ref{rlim}).
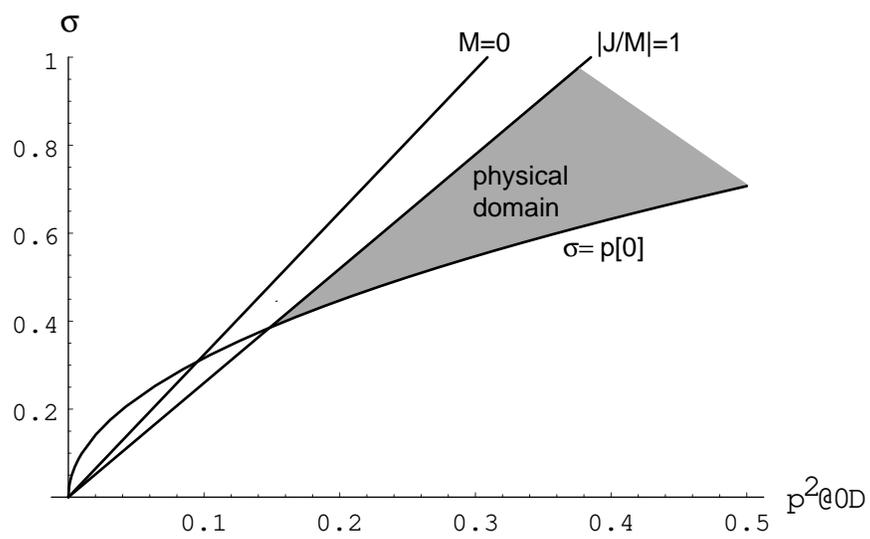
\item{Figure 3.} Plot, in the $(\sigma,\ p^2[0])$--plane, of the limiting curves $\sigma = p[0]$, $M=0$ (eq.
\ref{Mzero}) and
$-\Lambda\;J^2/M^2 =1$ (eq. \ref{JoverM}) (in units $|\Lambda| =1$), for  rotating, constant density, pressurized star models
with central angular velocity $|\omega|\ = 0.1\, \sqrt{-\Lambda}$. The gray area corresponds to the domain of physically allowed
values of
$\sigma$ and $p[0]$. 
\end{itemize}
\end{document}